\documentclass[twocolumn,showpacs,preprintnumbers,prd,superscriptaddress,nofootinbib]{revtex4-1}

\usepackage{amsmath}
\usepackage{amsfonts}
\usepackage{amssymb}
\usepackage{graphicx}
\usepackage{hyperref}
\usepackage{booktabs}
\usepackage{cleveref}
\usepackage{color}
\usepackage{mathrsfs} 
\usepackage{comment}

\Crefname{equation}{Eq.}{Eqs.}
\Crefname{figure}{Fig.}{Figs.}
\Crefname{section}{Sec.}{Secs.}
\Crefrangeformat{equation}{Eqs.~(#3#1#4)--(#5#2#6)}

\begin{document}

\title{Forecasting constraints on deviations from general relativity in $f(Q)$ gravity with standard sirens}

\author{Rocco D'Agostino}
\email{rocco.dagostino@unina.it}
\affiliation{Scuola Superiore Meridionale (SSM), Largo S. Marcellino 10, I-80138 Napoli, Italy.}
\affiliation{Istituto Nazionale di Fisica Nucleare (INFN), Sezione di Napoli, Via Cinthia 9, I-80126 Napoli, Italy.}

\author{Rafael C. Nunes}
\email{rafadcnunes@gmail.com}
\affiliation{Instituto de F\'{i}sica, Universidade Federal do Rio Grande do Sul, 91501-970 Porto Alegre RS, Brazil}
\affiliation{Divis\~ao de Astrof\'isica, Instituto Nacional de Pesquisas Espaciais, Avenida dos Astronautas 1758, S\~ao Jos\'e dos Campos, 12227-010, SP, Brazil}

\begin{abstract}
In this work, we explore how modified gravity theories based on the non-metricity scalar, known as $f(Q)$ gravity, affect the propagation of gravitational waves from inspiraling of binary systems. We discuss forecast constraints on $f(Q)$ gravity by considering standard siren events in two contexts: i) simulated sources of gravitational waves as black hole - neutron star binary systems, emitting in the frequency band of the third-generation detector represented by the Einstein Telescope (ET); ii) three standard siren mock catalogs based on the merger of massive black hole binaries that are expected to be observed in the operating frequency band of the Laser Interferometer Space Antenna (LISA). We find that, within the ET sensitivity, in combination with supernova and cosmic chronometer data, it will be possible to test deviations from general relativity at $<3\%$ accuracy in the redshift range $0<z<5$, while the main free parameter of the theory is globally constrained at 1.6\% accuracy within the same range. In light of LISA's forecasts, combined with supernova and cosmic chronometer data, in the best scenario, we find that the main free parameter of the theory will be constrained at 1.6\% accuracy up to high redshifts. Therefore, we conclude that future gravitational wave observations by ET and LISA  will provide a unique way to test, with good accuracy, the nature of gravity up to very large cosmic distances.
\end{abstract}


\maketitle
\preprint{ET-0236A-22}
\section{Introduction}

One of the greatest challenges in contemporary physics is to provide a suitable description of the nature of the dark sector of the Universe, namely, dark matter and dark energy (DE) \cite{Peebles:2002gy,Copeland:2006wr,DAgostino:2019wko}, which constitute together approximately 95\% of the energy density of the cosmic content.  The simplest possible explanation for DE, namely the cosmological constant $\Lambda$, relates its nature to the vacuum energy density. Due to its great success to explain the majority of the observations, the Lambda-Cold Dark Matter ($\Lambda$CDM) model is considered the standard model of cosmology.
Nonetheless, the cosmological constant leads to serious problems from the theoretical point of view \cite{Weinberg:1988cp,Padmanabhan:2002ji,DAgostino:2022fcx}. Alternatively to the $\Lambda$ term, one can consider extra degrees of freedom with a gravitational origin, \emph{i.e.}, arising from a gravitational modification that possesses general relativity (GR) as a particular limit. The modified gravity (MG) scenarios, in fact, may allow for extensions of the $\Lambda$CDM model and can drive the accelerated expansion of the Universe at late times, as well as explain various observations at the cosmological and astrophysical levels (see \cite{Clifton:2011jh, Ishak:2018his, Nojiri:2017ncd, CANTATA:2021ktz} for a review).

From an observational perspective, looking for new astrophysical sources, through a direct manifestation of gravitational effects, can provide rich physical information about the nature of gravity, which should play a key role to probe new (or rule out) MG or DE models. Gravitational wave (GW) astronomy provides an unprecedented opportunity to test gravitational physics in that direction. Currently, more than 90 coalescing compact binary events have already been observed during the three running stages of the LIGO/VIRGO mission \cite{LIGOScientific:2021djp}. One of the most promising prospects is the observation of  standard siren (SS) events \cite{Schutz:1986gp,Holz:2005df}. The latter are the GW analog of the astronomical standard candles and might be a powerful tool in view of constraining cosmological parameters through the information encoded in the luminosity distance provided by these events. To date, one event has been observed through a binary neutron star (BNS) merger at $z = 0.01$, namely the GW170817 event \cite{LIGOScientific:2017vwq, LIGOScientific:2017ync}. Preliminary cosmological information and the consequences of this observation are important to the understanding of our Universe locally. These observations were used to measure the Hubble constant \cite{LIGOScientific:2017adf} and also to impose strong constraints on MG/DE scenarios (see \cite{Kase:2018aps} for a review).

On the other hand, the detectability rate of the SS events from the current LIGO/VIRGO sensitivity is expected to be very low, as well as difficult to reach large cosmic distances. The central importance of GW astronomy is testified by the plans for the construction of several GW observatories in the future, such as the underground-based interferometers ET \cite{Maggiore:2019uih} and Cosmic Explore \cite{Reitze:2019iox}, and space-based interferometers such as LISA \cite{LISA:2017pwj}, DECIGO \cite{Kawamura:2020pcg} and TianQin \cite{TianQin:2015yph}, among others, to observe GWs in the most diverse frequency bands. The implications of cosmological studies using the SS have motivated focused studies on the nature of DE, MG, dark matter, and several other fundamental questions in modern cosmology \cite{Cai:2016sby, Du:2018tia, Zhang:2018byx, DAgostino:2019hvh, Yang:2019vni, Fu:2019oll, Cai:2017aea, Allahyari:2021enz, Belgacem:2017ihm, Nishizawa:2019rra, Bonilla:2019mbm, Odintsov:2022cbm, Cai:2021ooo, Matos:2021qne, Califano:2022cmo, Jiang:2021mpd, Pan:2021tpk, Tasinato:2021wol, Bonilla:2021dql, Mukherjee:2020mha, Kalomenopoulos:2020klp, Baker:2020apq, Mastrogiovanni:2020gua, Belgacem:2019zzu, Nunes:2019bjq, Califano:2022syd, Harry:2022zey, Ezquiaga:2021ler}. 

Looking through the geometrical character of gravity, it is pertinent to explore which equivalent manners gravity can be geometrized in. In fact, besides curvature, the other two fundamental quantities associated with the connection of a metric space are torsion and non-metricity \cite{BeltranJimenez:2019esp}. Among several viable candidates for MG theories, it has been proposed to construct scenarios where the gravitational interaction is mediated by non-metricity, while curvature and torsion are vanishing \cite{BeltranJimenez:2019esp,Bajardi:2020fxh,Ayuso:2020dcu,Capozziello:2022zzh}. These classes of models are known as $f(Q)$ gravity, where $Q$ is the non-metricity scalar. This approach could be important to describe gravity at a fundamental level because gravity can be dealt with as a gauge theory not requiring \emph{a priori} the validity of the Equivalence Principle. In the $f(Q)$ gravity context, the main dynamical equations in presence of matter have been derived in \cite{BeltranJimenez:2019tme}. From this study, modifications in the gravity sector emerge with respect to the $\Lambda$CDM model. Furthermore, observational constraints on the $f(Q)$ gravity have been performed using different observational probes for several parameterizations of the $f(Q)$ function \cite{Lazkoz:2019sjl,Frusciante:2021sio, Albuquerque:2022eac, Capozziello:2022wgl, Narawade:2022jeg, Ferreira:2022jcd, Khyllep:2022spx, Mandal:2021bpd, Atayde:2021pgb, Dimakis:2021gby, Anagnostopoulos:2021ydo, Zhao:2021zab, Khyllep:2021pcu, Solanki:2022ccf}.

The aim of this work is to obtain forecast constraints on $f(Q)$ gravity in light of three mock SS catalogs based on the merger of massive black hole binaries that are expected to be observed in the LISA operating frequency band, as well as from a mock SS catalog from black hole-neutron star mergers within the sensitivity predicted for the ET mission. In \cite{Ferreira:2022jcd}, a study was carried out to constrain the $f(Q)$ gravity through SS events. However, the present work differs from the previous one in two main aspects. Firstly, we here estimate deviations from GR by means of a different parameterization. Our choice, indeed, is based on a robust model-independent approach that minimizes possible \emph{a priori} biases towards a particular $f(Q)$ cosmological scenario. 
As a result, contrary to the aforementioned work, we do not assume a $\Lambda$CDM background evolution. Secondly, with regards to the ET perspective, we here use a mock catalog of black hole-neutron star mergers, from which we simulate detections up to redshift $z = 5$.

This paper is structured as follows. In \Cref{sec:theory}, we introduce the $f(Q)$ gravity framework and specify our theoretical setup. In \Cref{sec:data}, we present the datasets and the methodology used in our study. In \Cref{sec:results}, we show the results of our analysis and discuss the main physical consequences of our findings. Finally, in \Cref{sec:conclusions}, we outline our final considerations and perspectives.

\section{$f(Q)$ gravity and cosmology}
\label{sec:theory}

A fruitful way to obtain new hints on cosmic acceleration and, consequently, test the underlying gravitational theory, is to consider a different geometrical approach with respect to the Riemannian formulation. Specifically, in the present study, we shall explore the features of non-metricity at the cosmological level. 

For this purpose, we recall the most general form of the affine connection \cite{Jarv:2018bgs}:
\begin{equation}
\Gamma^{\lambda}_{\phantom{\alpha}\mu\nu} =
\left\lbrace {}^{\lambda}_{\phantom{\alpha}\mu\nu} \right\rbrace +
K^{\lambda}_{\phantom{\alpha}\mu\nu}+
 L^{\lambda}_{\phantom{\alpha}\mu\nu} \,,
\label{eq:connection}
\end{equation}
where $\left\lbrace {}^{\lambda}_{\phantom{\alpha}\mu\nu}\right\rbrace $ is the Levi-Civita connection: 
\begin{equation}
\left\lbrace {}^{\lambda}_{\phantom{\alpha}\mu\nu} \right\rbrace \equiv \dfrac{1}{2}\,g^{\lambda \beta} \left( \partial_{\mu} g_{\beta\nu} + \partial_{\nu} g_{\beta\mu} - \partial_{\beta} g_{\mu\nu} \right),
\end{equation}
with $g_{\mu\nu}$ being the metric tensor.
The last two terms of \Cref{eq:connection} are the contortion and disformation tensors, respectively:
\begin{align}
K^{\lambda}{}_{\mu\nu}& \equiv \frac{1}{2}\, g^{\lambda \beta} \left(\mathcal T_{\mu\beta\nu}+\mathcal T_{\nu\beta\mu} +\mathcal T_{\beta\mu\nu} \right), \\
L^{\lambda}{}_{\mu\nu}& \equiv \frac{1}{2}\, g^{\lambda \beta} \left( -Q_{\mu \beta\nu}-Q_{\nu \beta\mu}+Q_{\beta \mu \nu} \right)
\end{align}
where $\mathcal T^{\lambda}{}_{\mu\nu}\equiv \Gamma^{\lambda}{}_{\mu\nu}-\Gamma^{\lambda}{}_{\nu\mu}$ is the torsion tensor, while the non-metricity tensor reads 
\begin{equation}
Q_{\rho \mu \nu} \equiv \nabla_{\rho} g_{\mu\nu} = \partial_\rho g_{\mu\nu} - \Gamma^\beta{}_{\rho \mu} g_{\beta \nu} -  \Gamma^\beta{}_{\rho \nu} g_{\mu \beta}  \,.
\end{equation}
Therefore, the metric-affine spacetime is specified by the choice of the connection. In our study, we assume that geometry is provided by non-metricity, whereas torsion and curvature are both zero. Two independent traces can be associated with the non-metricity tensor depending on the contraction order, namely $Q_\mu=Q_{\mu\phantom{\alpha}\alpha}^{\phantom{\alpha}\alpha}$ and $\tilde{Q}^\mu={Q_{\alpha}}^{\mu\alpha}$. It follows that the non-metricity scalar can be expressed as \cite{BeltranJimenez:2017tkd}
\begin{equation}
Q=-\dfrac{1}{4}Q_{\alpha\beta\mu}Q^{\alpha\beta\mu}+\dfrac{1}{2}Q_{\alpha\beta\mu}Q^{\beta\mu\alpha}+\dfrac{1}{4}Q_{\alpha}Q^{\alpha}-\dfrac{1}{2}Q_{\alpha}\tilde{Q}^\alpha\,.
\end{equation} 

As for the cases of curvature-free or torsionless scenarios, one may consider theories of gravity that are based on a generic function of the non-metricity scalar, the so-called $f(Q)$ theories, whose action is given by\footnote{Here, we use units such that $c=1=8\pi G$.}
\begin{equation}
S=\int  d^4x\, \sqrt{-g}\left[\dfrac{1}{2}f(Q)+\mathcal{L}_m\right],
\label{eq:action}
\end{equation}
where $\mathcal{L}_m$ is the matter field Lagrangian, and $g$ is the determinant of $g_{\mu\nu}$.
Notice that, up to a total derivative, the above action and the Einstein-Hilbert one are equivalent for $f(Q)=Q$. Thus, GR is recovered as soon as the connections are globally vanishing and the non-metricity tensor can be written in terms of the metric only \cite{BeltranJimenez:2019esp,DAmbrosio:2021pnd}.

Varying action \ref{eq:action} with respect to the metric provides us with the field equations \cite{BeltranJimenez:2019tme}:
\begin{align}
&\dfrac{2}{\sqrt{-g}}\nabla_\alpha\bigg\{\sqrt{-g}\, g_{\beta\nu}\, f_Q\Big[-\dfrac{1}{2}L^{\alpha\mu\beta}- \dfrac{1}{8}\left(g^{\alpha\mu}Q^\beta+g^{\alpha\beta}Q^\mu\right) \nonumber \\
&+\dfrac{1}{4}g^{\mu\beta}(Q^\alpha-\tilde{Q}^\alpha)\Big]\bigg\}+f_Q\Big[-\dfrac{1}{2}L^{\mu\alpha\beta}-\dfrac{1}{8}\left(g^{\mu\alpha}Q^\beta+g^{\mu\beta}Q^\alpha\right)  \nonumber \\
&+\dfrac{1}{4}g^{\alpha\beta}(Q^\mu-\tilde{Q}^\mu)\Big]Q_{\nu\alpha\beta}+\dfrac{1}{2}{\delta^\mu}_\nu f={T^\mu}_\nu\,,
\label{eq:FE}
\end{align}
where $T_{\mu\nu}=-\frac{2}{\sqrt{-g}}\frac{\delta \sqrt{-g} \mathcal{L}_m}{\delta g^{\mu\nu}}$ is the energy-momentum tensor, and we have defined $f_Q\equiv \frac{\partial f}{\partial Q}$.

In order to analyze the cosmological features of $f(Q)$ gravity, let us consider the Friedman-Lema\^itre-Robertson-Walker (FLRW) metric with zero spatial curvature:
\begin{equation}
ds^2=-dt^2+a(t)^2\delta_{ij}dx^idx^j\,,
\end{equation}
where $a(t)$ is the scale factor as a function of the cosmic time $t$. To avoid trivial solutions that cannot go beyond GR, we assume the \emph{coincident gauge}  \cite{Hohmann:2021ast,DAmbrosio:2021zpm}, where the tangent space and spacetime share the same origin. Under this choice, the modified Friedmann equations take the form
\begin{align}
6H^2f_Q-\dfrac{1}{2}f&=\rho\,, \label{eq:first Friedmann} \\
\left(12H^2f_{QQ}+f_Q\right)\dot{H}&=-\dfrac{1}{2}(\rho+p)\,,  \label{eq:second Friedmann}
\end{align}
where $p$ and $\rho$ represent, respectively, the total pressure and density of the cosmic fluid. Furthermore, the non-metricity scalar is related to the Hubble parameter, $H\equiv \dot{a}/a$, through \cite{Capozziello:2022wgl}.
\begin{equation}
 Q=6H^2\,. 
 \label{eq:Q-H}
\end{equation}
As we focus our analysis on the late stages of the Universe's evolution, we can safely neglect the radiation contribution. Also, we assume that the cosmic fluid is totally made of pressureless matter, thus $p=0$ and 
\begin{equation}
    \rho=3H_0^2 \Omega_{m0}(1+z)^{3}\,,
    \label{eq:rho}
\end{equation}
where $z\equiv a^{-1}-1$ is the redshift, and $H_0$ and $\Omega_{m0}$ are the Hubble constant and the current matter density parameter, respectively\footnote{In our notation, the subscript `0' indicates the present-day values of the cosmological parameters, namely at $z=0$.}.

To work out the cosmic dynamics in $f(Q)$ gravity, one needs to specify the non-metricity function. A common approach is to assume \emph{a priori} the form of $f(Q)$ and then check for possible deviations from GR arising from the resulting dynamics. However, such a procedure may be affected by misleading conclusions due to possible biases inherent in the chosen model. 

Nevertheless, the aforementioned issues might be alleviated by resorting to the cosmographic method \cite{Visser:2004bf,Capozziello:2017nbu,Capozziello:2020ctn,Aviles:2012ay}, which has proven to be a powerful tool when applied to DE/MG scenarios \cite{Capozziello:2019cav,Mandal:2020buf,Capozziello:2018aba,Capozziello:2017uam,Capozziello:2017ddd}. In the specific case of $f(Q)$ gravity, we shall adopt the results obtained in the previous work \cite{Capozziello:2022wgl}, where the functional form of $f(Q)$ has been reconstructed by means of a  kinematic model-independent analysis on the background low-redshift measurements. Thus, in the present study, we consider the function
\begin{equation}
    f(Q)=\alpha+\beta Q^n\,,
    \label{eq:model}
\end{equation}
where $\alpha$, $\beta$ and $n$ are treated as free parameters. Besides being suggested directly from observations, this test function allows for a simple test of the deviations from GR ($\Lambda$CDM), which is recovered for $\beta=1=n$ and $\alpha=0$ $(\alpha>0)$.

The extra free parameters with respect to the $\Lambda$CDM model affect also the cosmological evolution at the perturbation level, as attested by the effective gravitational constant, $G_\text{eff}\equiv G/f_Q$ \cite{BeltranJimenez:2019tme}.
In particular, taking into account \Cref{eq:Q-H,eq:model}, we find
\begin{equation}
    \frac{G_\textbf{eff}(z)}{G}=\dfrac{(6H^2(z))^{1-n}}{n \beta}\,,
    \label{eq:Geff}
\end{equation}
for $\beta\neq 0\neq n$.
The effect induced by the effective gravitational constant on the GW propagation is measured through the quantity \cite{LISACosmologyWorkingGroup:2019mwx}
\begin{equation}
    d_\text{GW}(z)=\sqrt{\frac{G_\text{eff}(z)}{G_\text{eff}(0)}}\, d_L(z)\,,
    \label{eq:dGW}
\end{equation}
where $d_L(z)$ is the background luminosity distance, 
\begin{equation}
    d_L(z)=(1+z)\int_0^z \frac{dz'}{H(z')}\,.
    \label{eq:dL}
\end{equation}
Thus, in view of \Cref{eq:Geff}, from \Cref{eq:dGW} we obtain
\begin{equation}
    d_\text{GW}(z)= E(z)^{n-1}d_L(z)\,,
    \label{eq:dGW_1}
\end{equation}
where $E(z)\equiv H(z)/H_0$ is the dimensionless Hubble parameter. It is worth stressing that, as soon as $n=1=\beta$, $G_\text{eff}=G$ as in GR, and the GW propagation recovers the predictions of the $\Lambda$CDM model, characterized by
\begin{equation}
    E_{\Lambda\text{CDM}}(z)=\sqrt{\Omega_{m0}(1+z)^3+1-\Omega_{m0}}\,.
    \label{eq:E_LCDM}
\end{equation} 

\section{Datasets and Methodology}
\label{sec:data}

In light of the main scope of this work, we generate mock data inspired by the possibility of future observations of SS events. In particular, we are here interested in SS events to be detected by two different observatories, namely ET and LISA. We provide a brief description of our samples in the following.

\subsection{Einstein Telescope}

The ET is a third-generation ground-based detector, covering the frequency range $1 - 10^4$ Hz. The ET is expected to be ten times more sensitive than the current advanced ground-based detectors. We refer the reader to \cite{Maggiore:2019uih} for a presentation of the scientific objectives of the ET observatory. The ET conceptual design study predicts an order of $10^3$ - $10^7$ detections per year. After 10 years of operation, the ET is expected to detect $\sim$1000 GW SS events from the black hole-neutron star mergers up to $z=5$ \cite{Cai:2017aea}. 

Our goal, thus, is to generate a luminosity distance catalog matching the expected sensitivity of the ET after 10 years of operation. In particular, we generate 1000 triples ($z_i$, $d_L(z_i)$, $\sigma_i$), with $z_i$ being the redshift of the GW source, $d_L$ the measured luminosity distance, and $\sigma_i$ the uncertainty on the latter. There are three aspects to take into consideration in the mock data generation process: the fiducial cosmological model enters both in $z_i$ (or more precisely into the redshift distribution of expected sources) and $d_L$; the expected type of GW sources enter in $z_i$; finally, the instrumental and physical specifications enter in $\sigma_i$. In our case, we fix the fiducial model to the Planck-$\Lambda$CDM baseline parameters \cite{Planck:2018vyg}. 
The ET sensitivity we make use of in this work corresponds to the ET-D curve model\footnote{\url{https://www.et-gw.eu/index.php/etsensitivities}}, which include the most relevant fundamental noise contributions \cite{Hild:2010id}.

\begin{figure}
   \includegraphics[width=3.2in]{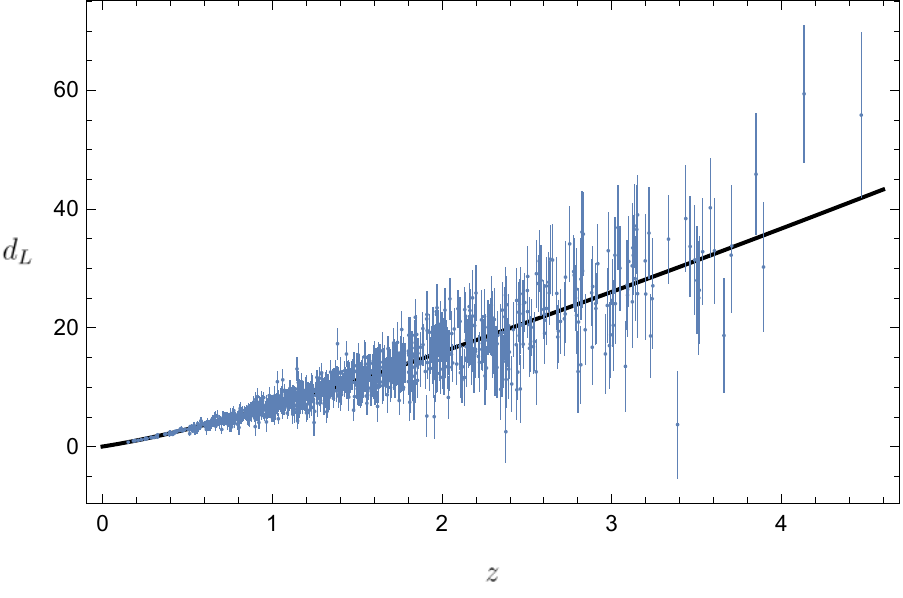}
    \caption{Simulated luminosity distance measurements with relative $1\sigma$ uncertainties from the mock ET catalog. The black curve refers to the best-fitted $\Lambda$CDM model.}.
    \label{fig:ET_LCDM}
\end{figure}

The whole methodology to generate the mock data is already very well-known and widely used in the literature. The features of this methodology are well described in previous works, such as \cite{Cai:2017aea,DAgostino:2019hvh}.
We display in Fig.~\ref{fig:ET_LCDM} the ET simulated $d_L(z)$ measurements along with the corresponding $\Lambda$CDM best fit (see Table \ref{tab:LCDM results}).

\begin{figure}
   \includegraphics[width=3.2in]{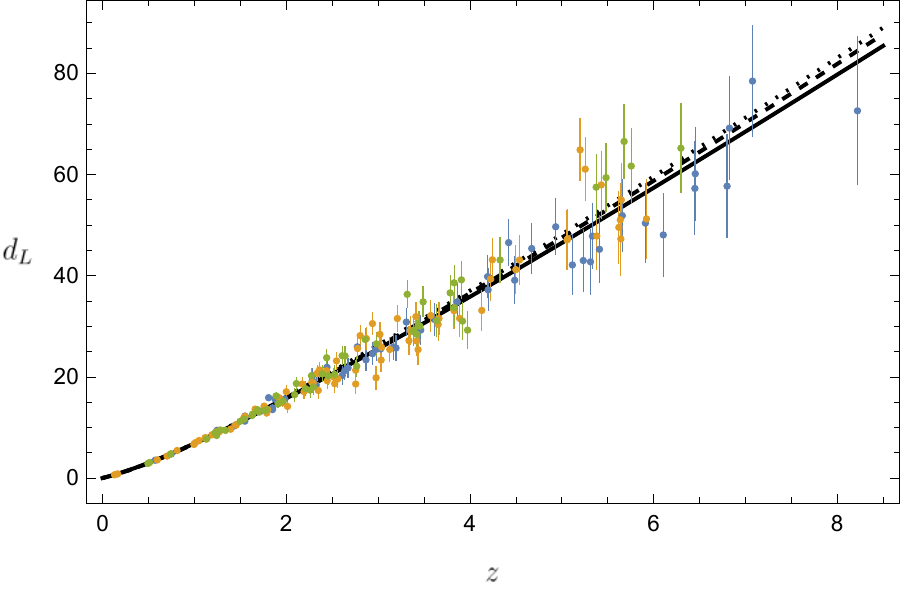}
    \caption{Simulated luminosity distance measurements with relative $1\sigma$ uncertainties from the mock LISA Delay (blue), No Delay (orange) and Pop III (green) catalogs. The black curves correspond to the $\Lambda$CDM best fits to LISA Delay (solid), No Delay (dashed) and Pop III (dotted) data.}
    \label{fig:LISA_LCDM}
\end{figure}

\subsection{LISA}

LISA will operate in the millihertz band with the objective to be an all-sky GW survey. Science with LISA brings opportunities and challenges in terms of complications arising from its motion around Earth. Basically, LISA can be thought of as two detectors,  and it will be launched in three identical drag-free spacecraft forming an equilateral triangle, with an arm length of about $2.5 \times 10^6$ km \cite{Cutler:1997ta}. 

Among astrophysical sources, LISA can reach Galactic binaries, stellar origin black hole binaries and extreme-mass-ratio inspirals \cite{Babak:2017tow}, and massive black hole binaries (MBHBs). See \cite{Seoane:2021kkk} for a presentation of the scientific objectives of the LISA mission. The most probable LISA sources with electromagnetic counterparts are MBHBs. In particular, MBHBs are supposed to merge in gas-rich environments and within the LISA frequency band, allowing for electromagnetic follow-ups to determine their $z$. Theoretical models and simulations can predict the redshift distribution and merger rate of MBHBs. Depending on the initial conditions for black hole formation at high $z$, there are two scenarios, namely, the light seed and the heavy seed ones.  In the light seed scenario, massive black holes are assumed to grow from the remnants of population III (pop III) stars forming at $z \in [15,20]$. In the heavy seed scenario, on the other hand, massive black holes are assumed to form from the collapse of protogalactic disks. The result of the scenarios produces three categories of population models named Pop III, Delay and No Delay \cite{Tamanini:2016zlh}. Our catalog is based on the model presented in \cite{Tamanini:2016zlh, Tamanini:2016uin}. The redshift distribution of MBHBs SS of our mock sample is displayed in Figs.~1 and 2 in \cite{Allahyari:2021enz}. 

In this case, we adopt the LISA sensitivity provided in \cite{Robson:2018ifk}, where the full sensitivity curve\footnote{\url{https://github.com/eXtremeGravityInstitute/LISA_Sensitivity}} is constructed by combining the galactic and the instrumental noises for a 4-year mission lifetime. Similar to the ET simulation, in the LISA mock generation data we fix the fiducial model to the Planck-$\Lambda$CDM baseline parameters \cite{Planck:2018vyg}.
In Fig.~\ref{fig:LISA_LCDM}, we show the simulated measurements of $d_L(z)$ from the all LISA catalogs with the corresponding $\Lambda$CDM best fits (see Table \ref{tab:LCDM results}).

\begin{table*}
\begin{center}
\setlength{\tabcolsep}{0.7em}
\renewcommand{\arraystretch}{2}
\begin{tabular}{c c c c c}
\hline
\hline
Dataset & $H_0$ & $\Omega_{m0}$ \\
\hline
ET & $ 67.69^{+0.63(1.26)}_{-0.63(1.23)} $ & $0.311_{-0.017(0.034)}^{+0.018(0.036)}$   \\
LISA\,(delay) & $ 64.42^{+1.38(2.71)}_{-1.41(2.91)}$ & $0.386^{+0.035(0.080)}_{-0.042(0.075)}$ \\
LISA\,(no delay) & $67.54^{+0.67(1.34)}_{-0.68(1.32)}$ & $0.317^{+0.017(0.035)}_{-0.017(0.033)}   $
\\
LISA\,(pop III) & $67.39^{+0.94(1.83)}_{-0.93(1.84)} $ & $0.306^{+0.023(0.046)}_{-0.023(0.042)}$
\\
SN\,+\,CC & $69.15_{-1.94(3.82)}^{+1.95(3.78)}$ &  $0.296_{-0.030(0.066)}^{+0.027(0.058)}$  \\
SN\,+\,CC\,+\,ET & $67.85^{+0.55(1.09)}_{-0.55(1.10)}$ & $0.307_{-0.015(0.028)}^{+0.015(0.030)}$ 
\\
SN\,+\,CC\,+\,LISA\,(delay) & $66.48^{+0.96(1.90)}_{-0.95(1.93)}$  & $0.333^{+0.021(0.043)}_{-0.022(0.040)}$
\\
SN\,+\,CC\,+\,LISA\,(no delay) & $67.77^{+0.62(1.21)}_{-0.62(1.22)}$ & $0.311^{+0.015(0.030)}_{-0.015(0.028)}   $
\\
SN\,+\,CC\,+\,LISA\,(pop III) &  $67.64^{+0.77(1.51)}_{-0.77(1.54)}$ & $0.301^{+0.017(0.036)}_{-0.017(0.033)}   $
\\
\hline
\hline
\end{tabular}
\caption{Summary of the MCMC results at the 68\% (95\%) c.l. for the $\Lambda$CDM model.}
\label{tab:LCDM results}
\end{center}
\end{table*}

\begin{table*}
\begin{center}
\setlength{\tabcolsep}{1em}
\renewcommand{\arraystretch}{2}
\begin{tabular}{c c c c c}
\hline
\hline
Dataset & $H_0$ & $\Omega_{m0}$ & $\beta$ & $n$\\
\hline
SN\,+\,CC & $68.59_{-2.69(5.46)}^{+2.69(5.18)}$& $0.386_{-0.144(0.279)}^{+0.148(0.260)}$  & $1.361_{-0.349(0.890)}^{+0.498(0.752)}$ & $0.993^{+0.022(0.044)}_{-0.022(0.042)}$  
\\
SN\,+\,CC\,+\,ET & $67.69_{-0.62(0.21)}^{+0.63(1.23)}$ & $0.315_{-0.151(0.246)}^{+0.150(0.249)}$ & $1.149_{-0.559(0.811)}^{+0.568(0.812)}$ &  $0.988_{-0.016(0.031)}^{+0.016(0.033)}$ 
\\
SN\,+\,CC\,+\,LISA\,(delay) & $ 66.35^{+1.16(2.26)}_{-1.17(2.22)}$ & $0.421_{-0.149(0.254)}^{+0.143(0.263)}$ & $1.307^{+0.448(0.731)}_{-0.393(0.770)}$ & $0.996^{+0.016(0.033)}_{-0.016(0.030)}$  \\
SN\,+\,CC\,+\,LISA\,(no delay) & $67.71^{+0.67(1.32)}_{-0.67(1.30)}$ &  $0.341^{+0.116(0.248)}_{-0.138(0.230)}$ & $1.132^{+0.387(0.731)}_{-0.408(0.713)}$ &  $0.995^{+0.015(0.030)}_{-0.015(0.029)}   $ \\
SN\,+\,CC\,+\,LISA\,(pop III) & $67.20^{+0.89(1.75)}_{-0.88(1.70)}$  & $0.386^{+0.137(0.254)}_{-0.137(0.241)}$ & $1.453^{+0.508(0.720)}_{-0.334(0.857)}$ &  $0.982^{+0.035(0.016)}_{-0.018(0.033)}   $ \\
\hline
\hline
\end{tabular}
\caption{Summary of the MCMC results at the 68\% (95\%) c.l. for the $f(Q)$ model under study. For $\beta=n=1$, we recover GR and the $\Lambda$CDM cosmological scenario.}
\label{tab:f(Q) results}
\end{center}
\end{table*}

\begin{figure*}
    \includegraphics[width=5in]{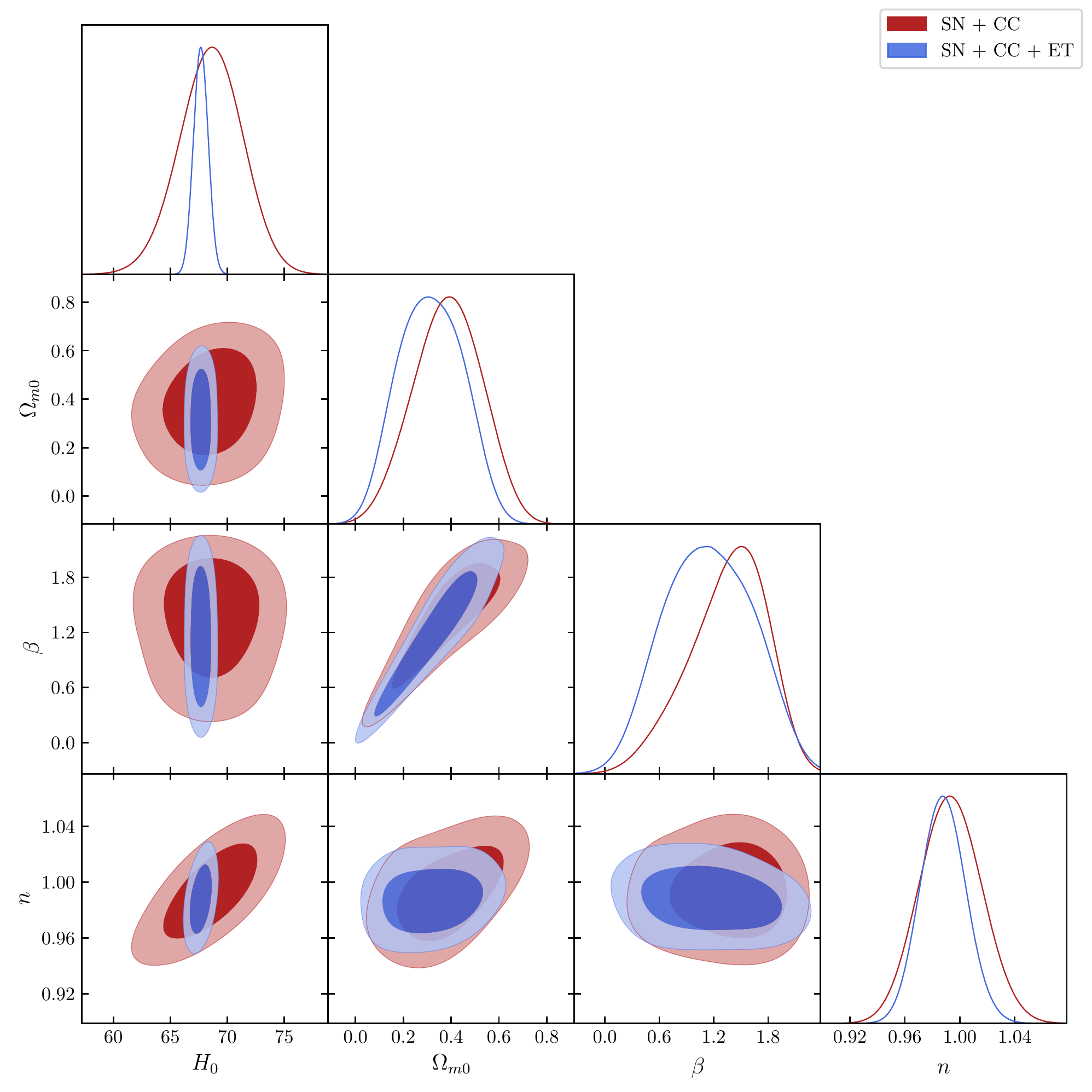}
    \caption{68\% and 95\% c.l. marginalized contours, with posterior distributions, as a result of the MCMC analysis using the ET mock data.}
    \label{fig:fQ_ET}
\end{figure*}

\begin{figure*}
   \includegraphics[width=5in]{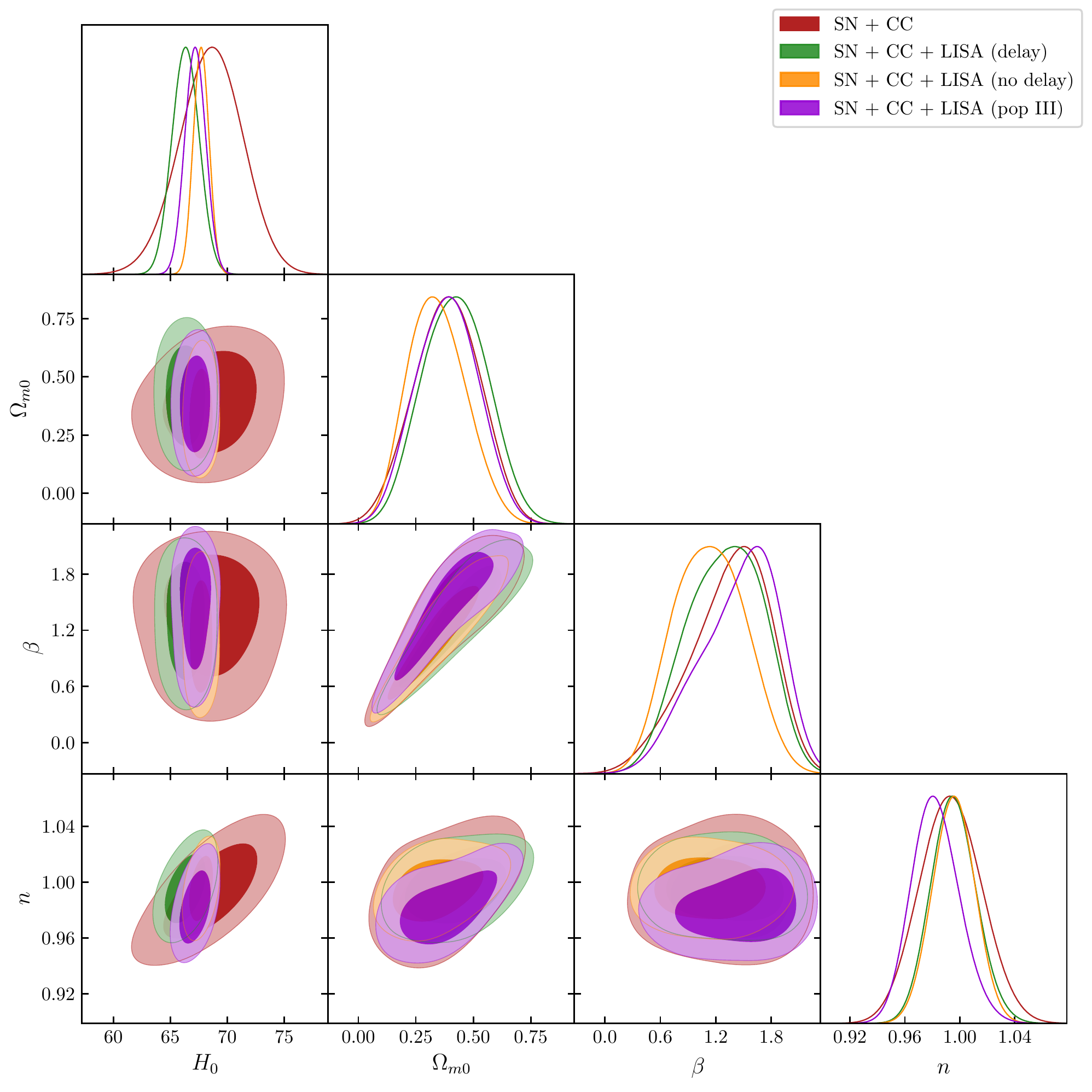}
    \caption{68\% and 95\% c.l. marginalized contours, with posterior distributions, as a result of the MCMC analysis using the LISA mock data.}
    \label{fig:fQ_LISA}
\end{figure*}

\section{Results and discussion}
\label{sec:results}

In this section, we present and discuss the results obtained from our numerical analysis of cosmological observations. In particular, to complement the GW SS simulated events from the ET and LISA experiments, 
we considered the low-redshift measurements of type Ia supernovae (SN) and cosmic chronometers (CC). We refer to \Cref{appendix} for the details on SN and CC datasets. 

\subsection{Monte Carlo analysis}

We test deviations from GR and the $\Lambda$CDM model by using the Markov Chain Monte Carlo (MCMC) method to analyze the $f(Q)$ model under consideration in this work.
In order to estimate observational constraints on the free parameters, we apply the Metropolis-Hastings algorithm \cite{Hastings70}, where the likelihood function for the GW SS mock dataset is built under the form
\begin{equation}
\mathscr{L}_\text{GW} \propto \exp \left\{ -\dfrac{1}{2} \sum_{i=1}^{N} \left[ \dfrac{d_{\text{GW},i}^{(obs)} - d_\text{GW}^{(th)}(z_i)}{\sigma_{d_{\text{GW},i}}}\right]^2 \right\},
\end{equation}
where $N$ is the size of the sample of each SS catalog. In the above equation, $d_{\text{GW},i}^{(obs)}$ are the simulated events with their associated uncertainties $\sigma_{d_\text{GW},i}$, while  $d_\text{GW}^{(th)}(z_i)$ is the theoretical prediction on each $i$th event. 

In a similar way, we build the likelihood functions for the SN and CC data (see \Cref{eq:L_SN,eq:L_CC}). As the latter are independent of the GW measurements, they may be combined with each other to obtain tighter constraints on the model parameters.  

To compare theoretical predictions and observational evidence, one needs to solve the modified Friedmann equations and find the cosmological dynamics. In our case, in view of \Cref{eq:model,eq:Q-H,eq:rho}, \Cref{eq:second Friedmann} becomes
\begin{equation}
6^{n-1} \beta  n (2 n-1) (z+1) H(z)^{2 n-1} H'(z)=\frac{3}{2} H_0^2 \Omega_{m0} (z+1)^3\,,
\label{eq:differential}
\end{equation}
where we have used the relation $\dot{H}=-(1+z)H(z) H'(z)$ to convert the time derivative into the derivative with respect to the redshift.
Thus, solving the first-order differential equation (\ref{eq:differential}) by means of the initial condition $H(z)=H_0$, we finally obtain
\begin{equation}
   H(z)= \left[H_0^{2 n}+\frac{H_0^2 \left[6^{1-n} \Omega_{m0} \left((z+1)^3-1\right)\right]}{\beta  (2 n-1)}\right]^{\frac{1}{2 n}},
   \label{eq:H(z)}
\end{equation}
for $\beta\neq 0$ and $n\neq 0,1/2$. The above solution can be then used to find the theoretical predictions for \Cref{eq:dGW_1} with the help of \Cref{eq:dL}. In the limit for $\beta\rightarrow 1$ and $n\rightarrow 1$, we recover the $\Lambda$CDM model as in \Cref{eq:E_LCDM}.

It is worth noticing that \Cref{eq:H(z)} does not involve the additive constant $\alpha$ of \Cref{eq:model}. 
This fact may be better understood by expressing the modified Friedmann equations  in light of the model \eqref{eq:model}. Specifically, from \Cref{eq:first Friedmann}, with the help of \Cref{eq:Q-H,eq:rho}, one finds
\begin{equation}
    \alpha +6 H_0^2 \Omega_{m0} (z+1)^3 =6^n \beta (2 n-1) H^{2n}\,,
\end{equation}
which, evaluated at the present time, provides
\begin{equation}
    \alpha = 6^n \beta  (2 n-1)H_0^{2 n}-6H_0^2\Omega_{m0}\,.
\end{equation}
Hence, the constant $\alpha$ does not represent a degree of freedom of our model, as it can always be expressed in terms of the other cosmological parameters. The physical meaning of $\alpha$ is easily revealed in the limit $n\rightarrow 1$ and $\beta \rightarrow 1$, when one obtains $H^2=H_0^2 \Omega_{m0}(1+z)^3+\alpha/6$. Then, recalling our hypothesis of a flat universe, we immediately can interpret $\alpha$ as the cosmological constant.

Therefore, the set of free parameters in our fitting procedure is $\theta=\{H_0,\Omega_{m0},\beta,n\}$. In particular, the estimates of $\beta$ and $n$ will quantify the deviations with respect to GR. 
In the realization of our MCMC analysis, the sampling is done by assuming the following uniform priors over $\theta$\footnote{In this paper, $H_0$ values are expressed in units of km/s/Mpc.}: 
\begin{subequations}
\begin{eqnarray}
    H_0 & \in & [50, 100] \,,\\
    \Omega_{m0} & \in & [0, 1]\,, \\
    \beta & \in & [-10,0) \cup (0,10]\,, \\
    n & \in & [-10,0) \cup (0,1/2) \cup (1/2,10] \,.
\end{eqnarray}
\end{subequations}

\noindent In what follows, we summarize our main results.

\subsection{Observational constraints}

Before proceeding to the forecast constraints on possible deviations from GR,  we summarize in Table~\ref{tab:LCDM results} the  results up to the $2\sigma$ confidence level (c.l.) from the statistical analyses of the $\Lambda$CDM model. First, we consider individually the four SS mock samples, namely, the ET sample and the LISA from the delay, no delay and pop III sample, respectively. As expected, given the total sample size (number of events), the accuracy on the free parameters, \emph{i.e.}, $H_0$ and $\Omega_{m0}$, is higher from either ET or LISA data with respect to the SN\,+\,CC measurements. In the latter case, we find 2.2\% accuracy on $H_0$, while 0.9\% accuracy from the ET analysis and 2.2\% from LISA (no delay) analysis. 
The analyses using the other LISA sources provide results with an intermediate accuracy with respect to the latter cases. Thus, on the one hand, the accuracy on $H_0$ that will be possible to achieve from SS events and, on the other hand, the fact that SS are independent of late-time probes such as SN, CC and BAO and have different systematic errors compared to the latter, clearly show that SS will
be an important complement in solving the $H_0$ tension\ in the future\footnote{See discussion in Section IX.7 in \cite{Abdalla:2022yfr} and references therein.}
Then, combining the SN\,+\,CC measurements with the SS mock events, we find that the accuracies on $H_0$ improve up to 0.8\%  using the ET forecasts, and 0.9\% using the LISA (no delay) forecasts. Thus, the SS events at very large cosmological distances to be observed in both the ET and LISA band can improve the current observational constraint in combination with other simple geometrical measurements. The same results apply to the $\Omega_{m0}$ parameter (c.f. Table~\ref{tab:LCDM results}). 
It is worth noticing that the results of the LISA (delay) sample are systematically different from those of the other two scenarios, which are roughly comparable to each other. In fact, LISA (delay) provides worse results in terms of accuracy due to the lower number of detectable SS, as also discussed in \cite{Tamanini:2016zlh}. Also, it is important to comment that the inclusion of high $z$ SS events, especially when their number density is low, may induce systematic effects in the cosmological analysis.

The main results concerning the statistical analyses for the $f(Q)$ gravity framework under consideration are summarized in Table \ref{tab:f(Q) results}. In this case, we do not report the results from GWs individually since they are not predictive enough. In fact, the MCMC constraints for the $f(Q)$ model are less stringent due to the presence of additional free parameters compared to the $\Lambda$CDM case. However, one can see the impact of considering the SS measurements from the comparison with the results based on SN\,+\,CC data only.
Due to the enlarged parameter space, the error bars will naturally increase compared to the $\Lambda$CDM model. When considering the SN\,+\,CC joint analysis, we find 4\% accuracy on $H_0$. However, from SN\,+\,CC\,+\,ET and SN\,+\,CC\,+\,LISA (no delay) data, we find 0.9\% and 1\% accuracy, respectively. Once again, the analyses using other LISA sources provide intermediary results to these accuracies. Thus, clearly, we can see that the addition of SS events will improve considerably the constraints on $H_0$ in the context of $f(Q)$ gravity.

Now, it is interesting to turn our attention to the parameters $\beta$ and $n$. In light of SN\,+\,CC data, we note 31\% and 2.2\% accuracy on $\beta$ and $n$, respectively. When considering the SS events, from CC\,+\,SN\,+\,ET data, we find 48\% and 1.6\%  accuracy on $\beta$ and $n$, respectively. From SN\,+\,CC\,+\,LISA (no delay), we find 34\% and 1.5\% accuracy on $\beta$ and $n$, respectively. It is worth to remark that the parameter space $\beta-n$ is statistically degenerate, despite the $\beta-n$ contours shows quite round shapes. In this regard, we note that the parameter $n$ is strongly correlated with both $H_0$ and $\Omega_{m0}$ when SN\,+\,CC are considered, while $\beta$ is only with $\Omega_{m0}$.

Here, the main parameter quantifying the model effects is $n$, which controls the power of gravitational correction to the GR prediction. We notice that the addition of SS events from both future experiments can improve the constraints on the minimal baseline, \emph{i.e.}, on the parameters $\Omega_{m0}$ and $H_0$. Apart from some statistical fluctuation, the final constraints on $\beta$ and $n$, are practically the same. Figures \Cref{fig:fQ_ET} and \ref{fig:fQ_LISA}  show the 2-dimensional parameter regions at 68\% and 95\% c.l. and the 1-dimensional posterior distributions for the $f(Q)$ model as results of the MCMC analysis of different combinations with SS data. 
Our results emerging from the SN\,+\,CC data analysis indicate no substantial evidence for deviations from GR, as the values of $n$ are consistent with the unity at the $1\sigma$ c.l.

\begin{figure*}
    \includegraphics[width=3.3in]{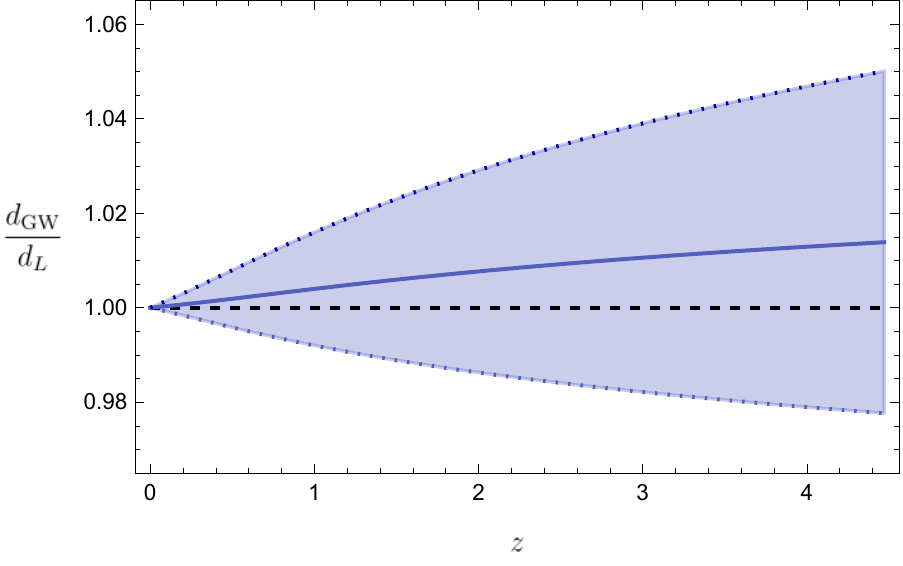} \qquad
     \includegraphics[width=3.3in]{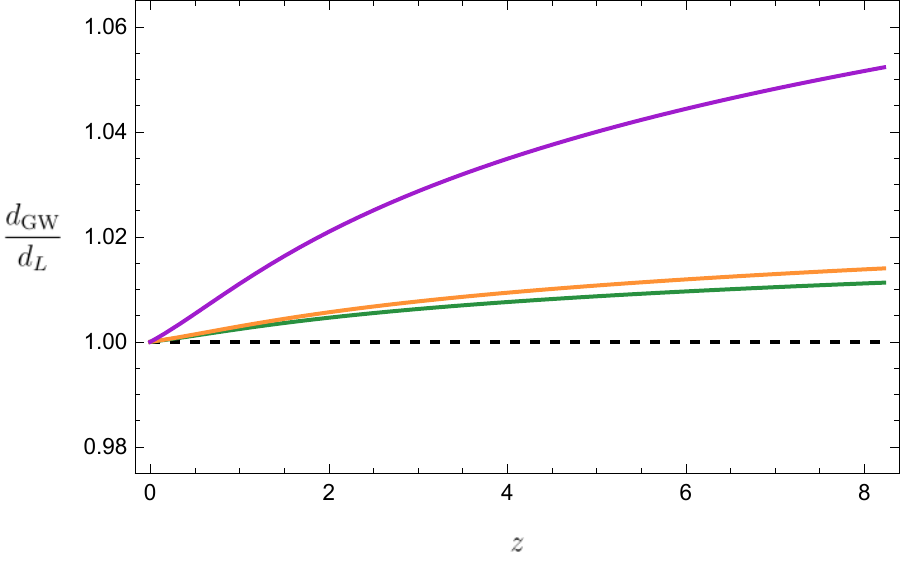}
     \caption{Effective luminosity distance for the $f(Q)$ model as a result of the MCMC analysis. \emph{Left panel.} The solid blue curve corresponds to the mean results from SN\,+\,CC\,+\,ET data, while the area between the dotted curves accounts for the relative $1\sigma$ uncertainties. \emph{Right panel.} The solid green, orange and violet curves correspond to the mean results from SN\,+\,CC\,+\,LISA\,(delay), SN\,+\,CC\,+\,LISA\,(no delay) and SN\,+\,CC\,+\,LISA\,(pop III)  data, respectively. The prediction of the $\Lambda$CDM paradigm is shown as a black dashed line.}
    \label{fig:dGW}
\end{figure*}

Furthermore, on the left panel of \Cref{fig:dGW}, we show a statistical reconstruction at the 1$\sigma$ c.l. of the effective luminosity distance, \Cref{eq:dGW}, under the perspective of the ET mock sample. We find an estimate of $d_{\rm GW}/d_L=1.01^{+0.03}_{-0.03}$ at $z \sim 4.5$, with gradually improving precision towards low $z$, as expected. This means that future measurements from ET will make it possible to test deviations from GR, under the $f(Q)$ gravity framework, at $\sim$3\% accuracy on $d_{\rm GW}/d_L$ ratio. 
Similarly, on the right panel of \Cref{fig:dGW}, we show the effective luminosity distance from the best-fit results using the LISA mock data. 

\section{Outlook and final remarks}
\label{sec:conclusions}

In this paper, we focused on the $f(Q)$ theories of gravity to test possible deviations from GR in light of  future GW detections. Specifically, taking into account the sensitivities of the ET and LISA experiments, we simulated mock SS events associated with black hole-neutron star binary systems and mergers of massive black hole binaries to probe the GW propagation in a FLRW Universe, where geometry is described by non-metricity.

Unlike previous approaches to $f(Q)$ gravity, our procedure relies on a robust model-independent method that minimizes possible biases induced by the choice of the underlying cosmology. For our purposes, we considered  a two-parameter extension of the $\Lambda$CDM model, where the power of the non-metricity scalar quantifies corrections with respect to Einstein's theory. In doing so, we worked out the cosmic dynamics at the background level, as well as at the perturbation level in terms of the effective gravitational constant of the theory.

After describing the methodology to generate mock SS measurements up to high redshifts from the perspective of the ET and LISA detectors, we presented the procedure to compare the observational evidence with the theoretical predictions.
In particular, a Monte Carlo numerical integration has been applied to constrain the free parameters of the model under consideration and test deviations with respect to the standard cosmological scenario. To improve the accuracy of our results, we complemented the simulated SS measurements with typical model-independent data at low redshifts. 

Our analysis shows that the inclusion of the SS measurements will considerably reduce the uncertainties on the $H_0$ estimate. More generally, adding the SS mock data up to large distances from both the ET and LISA missions will improve the accuracy of the whole parameter space. Besides, our study indicates no statistically significant deviations with respect to the GR predictions. 

Finally, adopting the results emerging from our joint analyses, we inferred the behavior of the effective luminosity distance up to very high redshifts. Specifically, when using the ET mock sample in combination with SN and CC data, we found that corrections to the standard luminosity distance could be tested at $\sim 3\%$ accuracy within the $f(Q)$ framework. On the other hand, no deviations bigger than 5\% are expected from the LISA perspective when combined with SN and CC measurements.

To conclude, the present study shows that future GW observations by the ET and LISA missions will offer a unique tool to test the nature of gravity up to very large cosmic distances with unprecedented precision.

\begin{acknowledgments}
R.D. acknowledges the support of Istituto Nazionale di Fisica Nucleare (INFN) - \textit{iniziativa specifica} QGSKY. 
The authors would like to thank Angelo Ricciardone and the Cosmology Division of the ET Observational Science Board (OSB) for the useful discussion on the manuscript.
The authors also thank the anonymous referee for his/her valuable comments and suggestions.
\end{acknowledgments}

\appendix 
\section{SN and CC datasets}
\label{appendix}

In this Appendix, we provide some details of the low-redshift cosmological observables\footnote{See also \cite{DAgostino:2018ngy,Bajardi:2022tzn}.} we use to complement the GW mock data in the statistical analysis on the $f(Q)$ model. 

The first complementary dataset we employ in our study is the Pantheon sample \cite{Pan-STARRS1:2017jku}, composed of 1048 SN Ia in the redshift range $0.01<z<2.3$. In this compilation, all the SN are standardized through the SALT2 light-curve fitter, in which the distance modulus is modelled as follows \cite{SNLS:2007cqk}:
\begin{equation}
\mu=m_B-M+\alpha x_1-\beta C + \Delta_M +\Delta_B ,
\label{mu_SALT2}
\end{equation}
where $m_b$ is the $B$-band apparent magnitude of each SN and $M$ is its
absolute magnitude, while $\Delta_M$ and $\Delta_B$ account for the host-mass galaxy and the distance bias corrections, respectively. Moreover, $x_1$ and $C$ are the stretch and color parameters of each SN light curve, respectively, with their relative coefficients $\alpha$ and $\beta$. On the other hand, the distance modulus predicted by a cosmological model is given as
\begin{equation}
\mu(z)=5\log_{10}\left[\dfrac{d_L(z)}{1\text{ Mpc}}\right]+25\,.
\end{equation}
As shown in \cite{Riess:2017lxs}, under the assumption of a flat universe, one can compress the full SN sample into a set of cosmological model-independent measurements of $E(z)^{-1}$. This approach allows us to properly marginalize over the SN nuisance parameters in the fitting procedure. Thus, taking into account the correlations among the $E^{-1}(z)$ measurements, we can write the likelihood function associated with the SN data as
\begin{equation}
\mathscr{L}_\text{SN}\propto \exp\left\{-\frac{1}{2}\mathbf{v}^\text{T} \mathbf{C}_\text{SN}^{-1} \mathbf{v}\right\} ,
\label{eq:L_SN}
\end{equation}
where $\mathbf{v}= E^{-1}_{obs,i}-E^{-1}_{th}(z_i)$ quantifies the difference between the measured values and the values predicted by a given cosmological model, and $\mathbf{C}_\text{SN}$ is the covariance matrix resulting from the correlation matrix given in \cite{Riess:2017lxs}. 
\\

The second complementary dataset is built upon the differential age approach developed in \cite{Jimenez:2001gg}, which represents a model-independent method to characterize the expansion of the Universe up to $z<2$.  In this technique, passively evolving red galaxies are used as cosmic chronometers (CC) to measure the age difference $(dt)$ of the universe at two close redshifts $(dz)$. Thus, one can estimate the Hubble parameter as
\begin{equation}
H(z)=-\dfrac{1}{(1+z)}\dfrac{dz}{dt} .
\end{equation}
In our analysis, we use the compilation of $H(z)$ uncorrelated measurements collected in \cite{Capozziello:2017buj} (see references therein). We can then write the likelihood function relative to the CC data as
\begin{equation}
\mathscr{L}_\text{CC}\propto\exp\left\{-\dfrac{1}{2}\sum_{i=1}^{N}\left[\dfrac{H_{obs,i}-H_{th}(z_i)}{\sigma_{H,i}}\right]^2\right\}  ,
\label{eq:L_CC}
\end{equation}
where $H_{obs,i}$ are the observed measurements with their relative uncertainties $\sigma_{H,i}$, while $H_{th}(z_i)$ are the theoretical values of the Hubble parameter obtained from using a specific cosmological model.

\end{document}